\documentclass[a4paper,10pt]{article}

\usepackage[utf8]{inputenc}
\usepackage[english]{babel}
\usepackage[margin=3.0cm]{geometry}
\usepackage{amsmath,amsfonts,amssymb}
\usepackage{fancyhdr}
\usepackage{physics}
\usepackage{graphicx}
\usepackage[framemethod=TikZ]{mdframed}
\usepackage{amsthm}
\usepackage{caption}
\usepackage{subcaption}
\usepackage{array,booktabs}
\usepackage{enumitem}
\usepackage{lipsum}
\usepackage{tensor}
\usepackage{cancel}
\usepackage{mathtools, nccmath}
\usepackage{empheq}
\usepackage{physics}
\usepackage[nottoc]{tocbibind}
\usepackage{lastpage}
\usepackage{hyperref}
\usepackage{cite}
\usepackage{tensor}
\usepackage{slashed}
\usepackage{mathrsfs}
\usepackage[vcentermath]{youngtab}
\usepackage{quiver}
\usepackage{cleveref}
\usepackage{ragged2e}
\usepackage{comment}
\usepackage{float}

\usepackage{MnSymbol}

\newcommand{\cN}{\mathcal{N}}

\newcommand{\be}{\begin{equation}}
\newcommand{\ee}{\end{equation}}
\newcommand{\bea}{\begin{eqnarray}}
\newcommand{\eea}{\end{eqnarray}}
\newcommand{\ba}{\begin{array}}
\newcommand{\ea}{\end{array}}

\newcommand{\cK}{\mathcal{K}}

\newcommand{\cM}{\mathcal{M}}

\def\repa{\raise4pt\hbox{$\square$}\mkern-14mu\raise-4pt\hbox{$\square$}}
\def\repab{\overline{\raise4pt\hbox{$\square$}\mkern-14mu\raise-4pt\hbox{$\square$}\mkern-1mu}}

\makeatletter \@addtoreset{equation}{section} \makeatother
\renewcommand{\theequation}{\thesection.\arabic{equation}}

\def\slashchar#1{\setbox0=\hbox{$#1$}           
   \dimen0=\wd0                                 
   \setbox1=\hbox{/} \dimen1=\wd1               
   \ifdim\dimen0>\dimen1                        
      \rlap{\hbox to \dimen0{\hfil/\hfil}}      
      #1                                        
   \else                                        
      \rlap{\hbox to \dimen1{\hfil$#1$\hfil}}   
      /                                         
   \fi}

\begin{document}

\begin{titlepage}

\begin{center}

\phantom{}
\vskip 1.5cm

{\Large \bf More on phase transitions 
 in  $\mathcal{N}=2$ massive gauge theories}

\vskip 1cm

{\bf Jeremías Aguilera-Damia\,${}^{a,c}$ and Jorge G.~Russo\,${}^{a,b}$} \\

\vskip 25pt

{\em $^a$  \hskip -.1truecm
\em Departament de F\' \i sica Cu\' antica i Astrof\'\i sica and Institut de Ci\`encies del Cosmos,\\ 
Universitat de Barcelona, Mart\'i Franqu\`es, 1, 08028
Barcelona, Spain.}

\vskip .4truecm

{\em $^b$  \hskip -.1truecm
\em Instituci\'o Catalana de Recerca i Estudis Avan\c{c}ats (ICREA),\\
Pg. Lluis Companys, 23, 08010 Barcelona, Spain.
 \vskip 5pt }

{\em $^c$  \hskip -.1truecm
\em Instituto de F\' \i sica La Plata,\\ Universidad Nacional de La Plata, C.C. 67, 1900 La Plata, Argentina}

\hskip 1cm

\noindent {\it e-mail:}  {\texttt jeremiasad@fqa.ub.edu , jorge.russo@icrea.cat}

\end{center}

\vskip 0.5cm
\begin{center} {\bf ABSTRACT}\\[3ex]
\end{center}

We study large $N$ phase transitions in $\mathcal{N}=2$  theories with gauge group $SU(N)$ and massive hypermultiplets in diverse representations.
Using supersymmetric localization  we identify cases where phase transitions occur.
In particular, we consider 
deformations of UV superconformal fixed points by giving two different masses to the
fundamental hypermultiplets, 
and show that these theories undergo a third-order phase transition at a critical value of the 't Hooft coupling.  This provides the first example of a phase transition driven by a mass deformation of a $\cN=2$ superconformal field theory.
We also comment on integrated correlators in these theories.

\noindent


\end{titlepage}


\renewcommand{\theequation}{\arabic{equation}}

\medskip

Superconformal gauge theories with $\mathcal{N}=2$ supersymmetry
and gauge group $SU(N)$ have been classified in \cite{Howe:1983wj,Koh:1983ir}
and there are essentially eight different cases, which are distinguished
by the matter content.
The partition function in all cases can be computed by supersymmetric localization \cite{Pestun:2007rz}. The explicit expressions of the partition functions for the eight different cases were given in \cite{Bourget:2018fhe} and
different aspects of these theories have been
extensively discussed
in many relevant papers \cite{Fiol:2015mrp,Billo:2019fbi,Beccaria:2020hgy,Fiol:2020ojn,Beccaria:2021hvt,Beccaria:2021vuc,Billo:2022xas,Bobev:2022grf}.

A notable feature of  these theories is that they have a
smooth dependence on the 't Hooft coupling $\lambda= g_{\rm YM}^2 N$ as $\lambda$ is increased from $\lambda=0$ to $\lambda=\infty$.
It was noticed in \cite{Russo:2013qaa,Russo:2013kea} that, in some $\cN =2$ theories, a radical change  occurs 
when the theories are deformed by mass terms.
A non-analytic behavior is exhibited by the free energy at specific values of the coupling, thereby signaling the presence of phase transitions.
Detailed studies of these phase transitions have been undertaken in \cite{Zarembo:2014ooa, Chen-Lin:2014dvz,Russo:2019lgq}, in which significant aspects of the  critical behavior have  been elucidated.

In  this note we will first extend the study of large $N$ phase transitions in massive $\mathcal{N}=2$ theories to other
superconformal field theories with gauge group $SU(N)$ within the classification 
of \cite{Koh:1983ir,Howe:1983wj}. We will  compute the vacuum expectation value (VEV) of the 1/2 supersymmetric Wilson loop and the free energy and examine their behavior near criticality. 

 Notably, in all previous examples  exhibiting non-trivial critical points, the undeformed theory ({\it i.e.} the UV fixed point) corresponds to either $\cN=4$ SYM or an asymptotically free theory. 
To date, no critical points have been identified in the context of massive deformations of  $\cN$=2 superconformal field theories. 
 In \cite{Russo:2013kea} different examples of mass  deformations of $\cN=2$ superconformal QCD were studied. All these models introduced a single mass scale parameter. It was found that, unlike analogous deformations in examples with non-trivial running, these theories do not exhibit phase transitions because the critical coupling in all these cases goes to infinity. A natural question is whether this feature is intrinsic of (mass deformed) $\cN=2$  superconformal field theories. By considering deformations with two mass scales, in this paper we will construct the first examples of mass deformed $\cN=2$ superconformal theories undergoing phase transitions at finite coupling. 

Finally, we will study some properties  of the derivatives of the free energy in massive SQCD with 
$N_f<2N$. 
\medskip

\paragraph{{\bf E} Theory (Rank-2 Antisymmetric+Symmetric):}
The simplest example corresponds to a deformation of $SU(N)$ SYM theory coupled to two hypermultiplets in the rank-2 symmetric and rank-2 antisymmetric representations. 
Following the ABCDE classification used in \cite{Beccaria:2020hgy}, 
we shall here refer to this theory as {\bf E}-theory. With this matter content, the theory is UV finite and represents  a supersymmetric conformal field theory. 
 Different studies of this theory using supersymmetric localization can be found in  \cite{Billo:2019fbi,Beccaria:2020hgy,Beccaria:2021hvt,Beccaria:2021vuc,Billo:2022xas,Bobev:2022grf}.

At leading order in the large $N$ expansion, there is a set of observables (including the free energy and the 1/2 BPS Wilson loop) that coincide with the corresponding ones in
$\cN=4$ SYM at leading order in the  large $N$ expansion.
 Correlation functions differ in the ``odd" sector. In the ``even" sector, the difference
 with $\cN=4$ SYM begins to show up to the order $1/N^2$.
This has been shown in \cite{Beccaria:2021vuc} by a calculation of  the coefficient of the $1/N^2$ term in the 1/2 BPS Wilson loop.
 
 We now deform the theory by assigning a mass $M$ and $-M$ respectively to the two hypermultiplets. The deformed partition function for the resulting {\bf E}$^*$ theory, computed by supersymmetric localization, is given by
\be
Z^{{\bf E}^*}=\int Da_i\  \frac{e^{-\frac{8\pi^2 N}{\lambda}\sum_i a_i^2}\ \prod_{i<j}(a_i-a_j)^2 H(a_i-a_j)^2}{\prod_i H(2a_i+M)\prod_{i<j}H(a_i+a_j+M)H(a_i+a_j-M)}\ ,
\ee
where 
\be
H(x)=\prod_{n=1}^\infty \left(1+\frac{x^2}{n^2}\right)^n e^{-\frac{x^2}{n} }=e^{-(1+\gamma) x^2}G(1+ix) G(1-ix)\ ,
\ee
where $G$ is the Barnes G-function and  $a_i$ are diagonal components of the VEV of the scalar field of the vector multiplet.
Here the radius of the four-sphere has been set to 1 (the dependence on the radius can be restored by $a\to aR$, $M\to MR$).

We are interested in performing the above integral in the large $N$ limit, where the sum over the eigenvalues $a_i$ is replaced by
an integral over $x$ with an eigenvalue density $\rho(x)$. Since the potential is symmetric under $x\to -x$, the minimal energy solution  is a symmetric distribution $\rho(x)=\rho(-x)$. Using this property, we arrive at the following saddle-point equation
\be\label{eq: N=2*}
\strokedint_{-\mu}^{\mu} dy\, \rho(y) \left[\frac{1}{x-y}-\cK(x-y)+\frac12\cK(x-y+M)+\frac12\cK(x-y-M)\right]=\frac{8\pi^2}{\lambda}x\ ,
\ee
where $\cK(x)=-H'(x)/H(x)$.
The saddle point equation \eqref{eq: N=2*} coincides with the one obtained in the study of $\cN=2^*$ SYM  \cite{Russo:2013qaa}.  $\cN=2^*$ SYM is obtained  as a massive deformation of $\cN=4$ $SU(N)$ SYM theory by giving mass to the adjoint $\cN=2$ hypermultiplet.
Its holographic description has been studied in \cite{Pilch:2000ue,Buchel:2000cn}, and some remarkable results were obtained for the theory on the four-sphere \cite{Bobev:2013cja}. 

Thus, to leading order in the large $N$ limit, the partition function of the  {\bf E}$^*$ theory coincides with the one of $\cN=2^*$ SYM. This implies that the planar equivalence of the free energy in {\bf E} theory and $\cN=4$ SYM extends to their corresponding $\cN=2$ massive deformations. As a direct consequence, in the decompactification limit at $R\to\infty$, the {\bf E}$^*$ theory presents an infinite sequence of phase transitions as  the coupling $\lambda$ is increased from $\lambda=0$ to $\lambda=\infty$ \cite{Russo:2013kea,Russo:2013qaa}. 
The first few critical couplings  are
$\lambda_c^{(1)} \approx 35.4, \ \lambda_c^{(2)} \approx 83,\ \lambda_c^{(3)} \approx 150$. Asymptotically, the critical couplings  approach the values  $\lambda_c^{(n)}\approx \pi^2 n^2$, $n\gg 1$.
The multiple phase transitions can be understood as arising from a resonance phenomenon, which occurs in the following manner. The eigenvalue density $\rho(x)$ has support 
 on a finite interval $[- \mu ,\mu ]$. The spectrum includes hypermultiplet states of masses $|a_i-a_j\pm M|$.
The width of the eigenvalue distribution increases with the coupling and, for a critical coupling,  one will have $2\mu=M$. At this coupling, new massless states (for which $a_i-a_j=\pm M$) will begin to contribute to the saddle point. Secondary resonances will appear whenever $2\mu =n M$, for any positive integer $n$, leading to new critical points.

\paragraph{C Theory ($N_f$ fundamentals + rank-2 Antisymmetric):} 
Another interesting $\cN=2$ superconformal gauge theory is obtained by adding $N_f=N+2$ fundamental hypermultiplets and a single rank-2 antisymmetric hypermultiplet. 
If $N_f<N+2$, the theory is asymptotically free, undergoes renomalization and dynamically generates a strong coupling scale $\Lambda $. Here we consider a  massive deformation, by giving mass $M$ to $N_f/2$ fundamentals and $-M$ to the remaining $N_f/2$ fundamentals, where we implicitly assume that $N_f$ is even.
The only reason for the symmetric assignment of masses is to simplify the solution for the eigenvalue density in terms of an even function $\rho(x)$.
For general $N_f<N+2$
the partition function is
\be
\label{cuatt}
Z^{N_f}=\int Da_i\  \frac{ e^{N(1-\zeta)(\log\Lambda+\gamma+1)\sum_i a_i^2}\ \prod_{i<j}(a_i-a_j)^2 H(a_i-a_j)^2}{\left[\prod_i H(a_i-M)H(a_i-M)\right]^{N_f/2}\prod_{i<j}H(a_i+a_j)}\ ,
\ee
where
\be
\zeta=\frac{N_f}{N+2}\ .
\ee
This partition function can be obtained by starting with the finite theory containing  $N+2$ fundamental hypermultiplets, where  $N+2-N_f$ fundamental hypermultiplets have mass $M_0$ (in the spirit of Pauli-Vilars regularization). The dynamically generated scale $\Lambda $ arises in  the limit of $M_0\to \infty$
\be\label{eq: Lambda}
\Lambda=M_0 e^{-\frac{8\pi^2}{\lambda(1-\zeta)}}\ . 
\ee
This procedure decouples the extra hypermultiplets and leaves a theory with $N_f<N+2$.

In the limit of $N\to \infty$, the partition function can be computed by a saddle point equation satisfied by the eigenvalue density
\be\label{eq: saddle point Nf<N}
2\strokedint_{-\mu}^{\mu} dy\, \rho(y) \left[\frac{2}{x-y}-\cK(x-y)\right]=-4(1-\zeta)(\log\Lambda+\gamma+1) x -\zeta\left[\cK(x+M)+\cK(x-M)\right]\ ,
\ee
with $\zeta=\lim_{N\to\infty}\frac{N_f}{N+2}$ fixed. 
We note that this equation is almost identical to the saddle-point equation that determines the partition function of
$\mathcal{N}=2$ theory coupled to $N_f<2N$ hypermultiplets of mass $M$, studied in \cite{Russo:2013kea}.
The only difference is a relative factor of 2 multiplying the term $1/(x-y)$ originating from the Vandermonde determinant.

This extends the observation by Fiol et al \cite{Fiol:2015mrp} to the non-conformal case.
This coincidence extends to
another  superconformal model in the classification of \cite{Howe:1983wj,Koh:1983ir},
 the {\bf B} theory,  obtained by adding $N_f=N-2$ fundamental hypermultiplets  together with a hypermultiplet in the rank-2 symmetric representation. In the large $N$ limit,
 the free energy of the  {\bf B} and {\bf C} models (with symmetric or antisymmetric representation and $N_f=N\pm 2$) coincide exactly.
  Like in the  model \eqref{cuatt},
 one may  consider a {\bf B}-type model with $N_f<N-2$ massive hypermultiplets and another  hypermultiplet in the symmetric representation.\footnote{Phase transitions in five-dimensional gauge theories with matter content similar  to the {\bf B} and {\bf C} models have been investigated in \cite{Santilli:2021qyt}.}.
 The corresponding  saddle-point equation at large $N$ is also described by \eqref{eq: saddle point Nf<N}.
 Therefore, in the large $N$ limit,   both {\bf B}$^*$ and {\bf C}$^*$ non-conformal models with $\zeta\leq 1$ have the same free energy.

We are interested in the dynamics of these theories in the decompactification limit, where all variables scale to large values, {\it i.e.} $(xR,\Lambda R, MR)\to \infty$. 
In this limit, one can use the asymptotic expansion
\be
\cK(x)=x\log x^2+2\gamma x+O(x^{-1})\ .
\ee
Within this regime, the first term inside the integral in \eqref{eq: saddle point Nf<N} (originating from the Vandermonde determinant) drops and  the saddle-point equation now becomes identical to the one corresponding to $N_f<2N$ SQCD, that is,  $SU(N)$ SYM coupled to $N_f/2$ fundamentals with masses $M$ and an equal number of fundamentals with mass $-M$. 

The saddle-point equation can be solved by differentiating  twice with respect to $x$. This yields
\be
\label{qzzz}
\strokedint_{-\mu}^\mu dy\frac{\rho(y)}{x-y} =\frac{\zeta}{2}\left(\frac{1}{x+M}+\frac{1}{x-M}\right)\ .
\ee
This equation has two different solutions depending on whether the points $x=\pm M$ are inside or outside the interval  $[-\mu,\mu]$:
\bea
&&\mu<M:\qquad\rho^w(x)= \frac{1-\zeta}{\pi\sqrt{\mu^2-x^2}} + \zeta \frac{M\sqrt{M^2-\mu^2}}{\pi \sqrt{\mu^2-x^2}(M^2-x^2)}\ ,
\label{mumenor}\\
\nonumber\\
&&\mu\geq M:\qquad \rho^s(x)= \frac{1-\zeta}{\pi\sqrt{\mu^2-x^2}} + \frac{\zeta}{2} \left(\delta(x+M)+\delta(x-M)\right)\ .
\label{mumayor}
\eea
The value $\mu=M$ is reached at the critical coupling $\Lambda_c=M/2$.
Thus the theory has two phases, a weak coupling phase,
with $\mu<M$ and $\Lambda<M/2$, and a strong coupling phase
with $\mu>M$ and $\Lambda>M/2$.
Further details will be given below in the more general context of the two-mass scale model.

\smallskip

The special case of  $\zeta=1$ is important because then the theory becomes finite (superconformal) in the UV. 
It was shown in \cite{Russo:2013kea} 
that the $\zeta=1$ massive theory does not undergo any phase transition. In this case, the width of the eigenvalue distribution is given by
\be
\mu =\frac{M}{\cosh \frac{4\pi^2}{\lambda}}\ .
\ee
It follows that $\mu < M$ for any finite coupling and the critical point $\mu=M$ is therefore never reached.
As a result, the theory presents a smooth behavior all the way from $\lambda=0$ to $\lambda=\infty$. 
In order to clarify whether phase transitions
can occur in more general mass deformations of superconformal theories,
we will now study the theory when the hypermultiplets have two different mass scales.

\bigskip

\paragraph{Deformation of $\cN=2$ SCFT's by two mass scales $m$ and $M$:} 
Let us now consider superconformal QCD where 
$\{N_f^m,\ N_f^m\}$ fundamental hypermultiplets have masses $\{m,\ -m\}$ and 
$\{N_f^M,\ N_f^M\}$ hypermultiplets have  masses $\{M,\ -M\}$, with  $m<M$. In addition, we assume that $N_f^{\rm tot}=2N_f^m+2N_f^M=2N$,
so that the theory is superconformal in the UV. 
We define the Veneziano parameters
$\zeta_m=N_f^m/N $ and $\zeta_M=N_f^M/N$ satisfying
\be
\zeta_m+\zeta_M =1\ .
\ee
At large $N$, the partition function is determined by a saddle-point.
The saddle point equation is
\bea
\strokedint_{-\mu}^\mu dy\, \rho(y)\left(\frac{1}{x-y}-\nu K(x-y)\right)&=& \frac{8\pi^2}{\lambda}x -\frac{\nu \zeta_m}{2}\left(K(x+m)+K(x-m)\right)
\nonumber\\
&-&\frac{\nu \zeta_M}{2}\left(K(x+M)+K(x-M)\right)\ .
\eea
Here $\nu =1$ describes the case of superconformal QCD, but the
same equation with $\nu=\frac12 $  describes
two other cases:

\medskip

\noindent a) 
 The deformation of $SU(N)$ SYM coupled to $N_f=N+2$ fundamental and a rank-2 antisymmetric hypermultiplets
 with $N_f^m/2$ ($N_f^M/2$) fundamentals of mass $m$ ($M$) and $N_f^m/2$ ($N_f^M/2$) fundamentals of mass $-m$ ($-M$).

\medskip

\noindent  b) The theory with $N_f=N-2$ fundamental and
rank-2 symmetric hypermultiplet, with the same mass deformation
(at large $N$, $N_f/N\to 1$).

\medskip

A  study of the  phases of a two-scale model with $N_f^{\rm tot}<2N$ fundamental hypermultiplets was carried out in the appendix of \cite{Russo:2017ngf}. Nevertheless, the deformation of the superconformal fixed point with $N_f^{\rm tot}=2N$ constitutes an exceptional case that was not addressed in the aforementioned paper.
In particular, in this case the gauge coupling  $\lambda$ does not run, unlike the $N_f^{\rm tot}<2N$ case, which has a dynamically generated scale $\Lambda_{\rm QCD}$. 
The present study will fill this gap and also clarify why the one-scale model has only one phase, whereas the theory can have multiple phase transitions if matter is added with different masses.

In the decompactification limit $R\to \infty$, the saddle point equation takes the following form
\be\label{xxx}
\nu\int_{-\mu}^\mu dy\, \rho(y)(x-y)\log(x-y)^2 = -\frac{8\pi^2}{\lambda}x +\frac12 \nu\zeta_mx\log (x^2-m^2)^2 +\frac12 \nu \zeta_M x\log (x^2-M^2)^2\ .
\ee
To solve this equation, it is advantageous to calculate the first and second derivatives of the equation with respect to the variable $x$. To leading order in the decompactification limit, for the first derivative one obtains 
\be\label{yyy}
\nu\int_{-\mu}^\mu dy\,\rho(y)\log(x-y)^2 = -\frac{8\pi^2}{\lambda}+ \frac{\nu\zeta_m}{2}\log (x^2-m^2)^2 +\frac{\nu\zeta_M}{2}\log (x^2-M^2)^2\ .
\ee
Differentiating one more time yields
\be
\label{zzz}
\strokedint_{-\mu}^\mu \frac{\rho(y)}{x-y} = \frac{\zeta_m}{2}\left(\frac{1}{x+m}+\frac{1}{x-m}\right) +\frac{\zeta_M}{2}\left(\frac{1}{x+M}+\frac{1}{x-M}\right)\ .
\ee
Similarly to the case of \eqref{qzzz}, the solution depends on whether the points $x=\pm m$ and $x=\pm M$ lie inside or outside
the cut $[-\mu,\mu]$.

\paragraph{Weak coupling phase ($\mu<m<M$):} 
When the coupling $\lambda$ is weak, the confining force in \eqref{xxx} is strong, and $\mu $  is pushed to small values.
As long as $\mu < m$, the poles at $x=\pm m$ and $x=\pm M$ in \eqref{zzz} lie outside the cut.
Under these circumstances, the unique normalized solution to \eqref{zzz} is given by
\be\label{eq: 2 scales weak solution}
\rho^w(x)=\zeta_m\frac{m\sqrt{m^2-\mu^2}}{\pi \sqrt{\mu^2-x^2}(m^2-x^2)}+\zeta_M\frac{M\sqrt{M^2-\mu^2}}{\pi  \sqrt{\mu^2-x^2}(M^2-x^2)}\ ,\qquad \mu< m\ .
\ee
The requirement that this solution  satisfies \eqref{xxx} gives rise to a transcendental equation for $\mu$,
\be\label{mueq}
\frac{\mu}{m+\sqrt{m^2-\mu^2}} \left(\frac{m+\sqrt{m^2-\mu^2}}{M+\sqrt{M^2-\mu^2}}\right)^{\zeta_M}=e^{-\frac{4\pi^2}{\lambda \nu}}\ ,
\ee
where we used $\zeta_m=1-\zeta_M$. This defines $\mu $ as a function of $\{ m,\, M,\, \zeta_M,\, \lambda \}$.

As $\lambda $ is increased, $\mu $ increases and a critical coupling is encountered at the point where
\be
\mu(\lambda_c)=m\ .
\ee
From \eqref{mueq}, we find
\be\label{lambdacr}
\lambda_c=\frac{4\pi^2}{\nu\zeta_M}\ \frac{1}{{\rm arccosh}\left(\frac{M}{m}\right)}\ .
\ee
 Notice that $\lambda_c\to \infty$  as $m\to M$, signaling the {\it absence} of phase transitions for the theory with a single-mass deformation.
 In this case, when $\nu=1$, the equations describe 
  SQCD with $N_f=2N$ hypermultiplets of mass $M$. 
 The same holds true for the case of $\nu=\tfrac{1}{2}$, describing the {\bf B}$^*$ and {\bf C}$^*$ theories with
 $\lim_{N\to\infty} \frac{N_f}{N}\to 1$ with a {\it single} mass $M$: these theories do not 
 undergo phase transitions.

\paragraph{Strong coupling phase ($m<\mu<M$):}
As  $\lambda$ is increased, the confining force becomes weaker  and the size of the cut becomes larger. For $\lambda>\lambda_c$, one has $\mu>m$. Assuming that $m<\mu<M$, the solution to \eqref{yyy} is now
\be
\rho^s(x)=\zeta_M\frac{M\sqrt{M^2-\mu^2}}{\pi \sqrt{\mu^2-x^2}(M^2-x^2)} +\frac{\zeta_m}{2}\left(\delta(x+m)+\delta(x-m)\right)\ .
\ee
The width of the eigenvalue distribution is obtained imposing  \eqref{xxx}. This yields
\be
\label{rrt}
\mu=\frac{M}{\cosh{\frac{4\pi^2}{\lambda\nu\zeta_M}}}\ .
\ee
For $\lambda\to\lambda_c$, one has $\mu\to m$, which shows that $\mu$ is continuous at the critical point.

As $\lambda $ is further increased, $\mu $ increases and asymptotically reaches the value $\mu\to M$ when $\lambda\to\infty$.  The poles at $x=\pm M$
lie outside the cut for any $0<\lambda <\infty$.
Consequently, there is no phase with $\mu>M$.

\paragraph{Critical behavior:}
The determination of the order of phase transition is achieved by studying the analytic properties of the free energy, 
$$
F=-\frac{1}{N^2}\log Z\ .
$$
The phase transition will manifest itself as a discontinuity in the $n$th-derivative of the free energy with respect to the coupling $\lambda $. Equivalently, we can study continuity
properties of the derivatives of $F$ with respect
to an inverse coupling defined as 
$$
\alpha \equiv \frac{8\pi^2}{\lambda}\ . 
$$
The first derivative of the free energy with respect to $\alpha $ is
\be
\frac{\partial F}{\partial\alpha }=\ \langle x^2\rangle\ ,
\ee
where
\be
\langle x^2\rangle =\int_{-\mu}^\mu dx\, \rho(x) \, x^2\ .
\ee
Let us denote as $F_w$ and $F_s$ the free energies in the weak $(\lambda<\lambda_c )$ and strong $(\lambda>\lambda_c )$ coupling regimes, respectively.
Computing $\langle x^2\rangle $ using the eigenvalue densities in the weak and strong coupling regimes, we find
\bea
&&\frac{\partial F_w}{\partial\alpha }=
\zeta_m\left(m^2-m\sqrt{m^2-\mu^2}\right)+\zeta_M\left(M^2-M\sqrt{M^2-\mu^2}\right)\ ,
\label{freedebil}\\
&&\frac{\partial F_s}{\partial\alpha }=
\zeta_m m^2+\zeta_M\left(M^2-M\sqrt{M^2-\mu^2}\right)\ .
\label{freestrong}
\eea
In the strong coupling phase $\mu $ is given by the simple formula \eqref{rrt}. Using this, we find
\be
\label{freeF}
F_s=\zeta_m m^2\alpha -2\nu\zeta_M^2  M^2 \log\left(1+e^{-\frac{\alpha}{\nu \zeta_M }}\right)+{\rm const.}
\ee
In the particular case $m=0$, only the strong coupling phase exists
and the theory (with $\nu=1$) becomes the ``partially massless" theory studied in \cite{Russo:2013kea}.
Setting $m=0$ in \eqref{freeF}, the resulting expression matches (4.32) of \cite{Russo:2013kea},  representing SQCD with $1-\zeta_M$ massless flavors and $\zeta_M$ flavors of mass $M$.

\medskip

From \eqref{freedebil} and \eqref{freestrong} it follows that the first derivative of the free energy is continuous at $\lambda=\lambda_c$, where $\mu(\lambda_c)=m$. 
The second derivative can be computed by using
the chain rule.
One obtains a strikingly simple result:
\bea
&&\frac{\partial^2 F_w}{\partial\alpha^2 }=\frac{1}{(\partial\alpha/\partial\mu )}\frac{\partial }{\partial\mu }\frac{\partial F_w}{\partial\alpha}=-\frac{\mu^2}{2\nu}\ ,
\nonumber\\
&&\frac{\partial^2 F_s}{\partial\alpha^2 }=\frac{1}{(\partial\alpha/\partial\mu )}\frac{\partial }{\partial\mu }\frac{\partial F_s}{\partial\alpha }=-\frac{\mu^2}{2\nu}\ .
\label{hesta}
\eea
Here we have used the explicit expressions for $\alpha(\mu)$ in the subcritical and supercritical regimes given in \eqref{mueq} and \eqref{rrt}.
Thus the second derivative is also continuous at $\mu=m$.
Finally, computing the third derivatives we obtain
\bea
&&\frac{\partial^3 F_w}{\partial\alpha^3 }=\frac{ \mu ^2}{2\nu^2}\frac{\left(\mu ^2-A \right) \left(\mu ^2-B \right)}{A \left(B-\mu ^2 \right)+\mu ^2 \zeta _M (A -B )}\ ,
\nonumber\\
&&\frac{\partial^3 F_s}{\partial\alpha^3 }=\frac{ \mu ^2 \sqrt{M^2-\mu ^2}}{2M \nu^2\zeta _M}\ ,
\eea
where we have defined
\be \nonumber
A\equiv m^2+m\sqrt{m^2-\mu ^2}\ ,\qquad B\equiv M^2+M\sqrt{M^2-\mu ^2}\ .
\ee
At the critical point, $\mu\to m$ and
\bea\label{disc third}
&&\frac{\partial^3 F_w}{\partial\alpha^3 }\to \frac{ m\sqrt{2m} \sqrt{m-\mu}}{2\nu^2(1-\zeta _M)}+O(m-\mu)\to 0\ ,
\nonumber\\
&&\frac{\partial^3 F_s}{\partial\alpha^3 }\to \frac{m ^2 \sqrt{M^2-m^2}}{2M \nu^2\zeta _M}\ .
\eea
Thus the third derivative of the free energy is discontinuous: the theory undergoes a  third-order phase transition.

The vacuum expectation value of the circular Wilson loop operator is given by
\be
\langle W\rangle =\int_{-\mu}^\mu dx \, \rho(x)\, e^{2\pi x}\ .
\ee
In the decompactification limit $\mu $ scales to infinity and one has the simple relation
$\langle W\rangle \to e^{2\pi\mu}$, or
\be
\log\langle W\rangle =2\pi\mu\ .
\ee

The Wilson loop is continuous but it presents a discontinuity in the first derivative
since 
\be
\frac{\partial \mu_w}{\partial \alpha}\bigg|_{\lambda\to\lambda_c^-}=0\ ,\qquad
\frac{\partial \mu_s}{\partial \alpha}\bigg|_{\lambda\to\lambda_c^+}=-\frac{m \sqrt{M^2-m ^2}}{M \zeta _M}\ .
\ee
\medskip

\subsubsection*{The case $\zeta_M=\zeta_m=1/2$}

Some important simplifications occur when $\zeta_M=\zeta_m=1/2$.
In this case the equation \eqref{mueq} for $\mu $
in the weak coupling regime can be solved explicitly. The solution is
\be
\mu^2=\frac{2M m\cosh(\alpha/\nu)-M^2-m^2}{\sinh^2(\alpha/\nu)}\ .
\ee
In this phase, $\mu\leq m $, as can also be seen by direct calculation:
\be
\label{masm}
\mu^2-m^2 = -\frac{(M-m\cosh(\alpha/\nu))^2}{\sinh^2(\alpha/\nu)}\leq 0\ .
\ee

Having an explicit expression for $\mu$ in both phases, we can now write down formulas for the free energy in explicit form, where the coupling dependence is manifest.
We find
\bea
&&\frac{\partial F_w}{\partial\alpha}=m M \text{csch}\left(\frac{\alpha }{\nu }\right)-\frac{1}{2}
   \left(m^2+M^2\right) \left(\coth \left(\frac{\alpha }{\nu
   }\right)-1\right)\ ,
\nonumber\\
&&\frac{\partial F_s}{\partial\alpha}=\frac{1}{2} \left(m^2+M^2-M^2 \tanh \left(\frac{\alpha }{\nu
   }\right)\right)\ .
   \label{finalx2}
\eea
The above formulas  include, as  particular cases, different models:
(a) the ``partially massless" $m=0$ case, discussed above; b) the massive deformation  of SQCD with $M=m$; c) the asymptotically free theory obtained by taking $M=\infty$ and integrating out $\zeta_M N$ fundamental hypermultiplets.

When $m=M$, the regime with $m<\mu<M$ disappears and as a result only the weak coupling phase remains.
In this case, after some simplifications, we find
\be
F_w=
-2\nu   M^2\log 
\left(1+e^{-\frac{\alpha }{\nu }}\right)+{\rm const.}\ 
\ee
Setting $\nu=1$, this agrees with (4.24) in \cite{Russo:2013kea}, representing  ${\cal N}=2$  SQCD with $N_f=2N$ fundamental hypermultiplets of mass $M$.

Finally, let us consider the asymptotically free SQCD theory obtained by taking the limit  $M=\infty$ with $\Lambda =M e^{-\frac{\alpha}{2\nu \zeta_M}}$ fixed.
For $\zeta_M=\tfrac{1}{2}$, in this limit we find
\bea
\label{x2debil}
&&\langle x^2\rangle_w=\frac{\partial F_w}{\partial\alpha}=2m\Lambda-\Lambda^2\ ,
\\
&&\langle x^2\rangle_s=\frac{\partial F_s}{\partial\alpha}=\frac{m^2}{2}+\Lambda^2\ ,\qquad 
\Lambda\equiv M e^{-\frac{8\pi^2}{\nu \lambda}}\ .
\eea
This reproduces the formulas (5.23) and (5.24) 
given in \cite{Russo:2013kea}. Note that the formula also holds for the $\nu=\tfrac12$ case corresponding to ${\bf B}^*$
and ${\bf C}^*$ theories with $\zeta_M=\tfrac12$.

\medskip

\paragraph{Weak coupling expansion:}
The weak coupling expansion can  be obtained explicitly for any value of $\zeta_M$, where $0<\zeta_M<1$, by solving the equation \eqref{mueq} for $\mu$ in perturbation theory.

For $\lambda \ll 1$, there is a large scale separation between the masses of the hypermultiplets and the typical scale $\Lambda_{\rm eff}$ governing the dynamics of the light degrees of freedom.
In this regime, integrating out the heavy fields leads to pure ${\cal N}=2$ SYM as the low energy effective fied theory, with the dynamical scale given by  
\be
\label{lamef}
\Lambda_{\rm eff}=\cM \ e^{-\frac{4\pi^2}{\nu \lambda}} \quad , \quad \cM = m^{\zeta_m}M^{\zeta_M}\ .
\ee
The resulting weak field expansion may be viewed as
an OPE in the effective field theory. We obtain
\be
\mu^2 =4 \Lambda_{\rm eff}^2-8\Lambda_{\rm eff}^4
   \left(\frac{\zeta_m}{m^2}+\frac{\zeta_M}{M^2}\right) 
   +12 \Lambda_{\rm eff}^6 \left(\left(\zeta _M-\zeta _m\right)\left(\frac{\zeta _M }{M^4}-\frac{\zeta _m }{m^4}\right)+\frac{4 \zeta _m \zeta _M}{m^2 M^2}\right)+O\left(\Lambda_{\rm eff}^8\right)\ ,
   \ee
and
\bea
 F_w&=&-2\nu \Lambda_{\rm eff}^2-\nu\Lambda_{\rm eff}^4 \left( \left(\frac{1}{m^2}-\frac{1}{M^2}\right) \zeta
   _M-\frac{1}{m^2}\right)
   \nonumber\\
    &-&\frac{2\nu  \Lambda_{\rm eff}^6}{3m^4 M^4} \left(\zeta _M \left(2
   \left(m^2-M^2\right)^2 \zeta _M+4 m^2 M^2-m^4-3
M^4\right)+M^4\right)+O\left(\Lambda_{\rm eff}^8\right)\ .
\label{opex}
\eea
Note that the  original expressions \eqref{freedebil}, \eqref{mueq} provide a resummation of the OPE expansion in  closed form.

A dramatic simplification occurs when $\zeta_m=\zeta_M=\frac12$. In this case one obtains
\be
\label{factoriz}
\frac{\partial F_w}{\partial\alpha}=
e^{-\frac{8\pi^2}{\nu \lambda}}
\left( 2mM -(m^2+M^2)e^{-\frac{8\pi^2}{\nu \lambda}}\right)\sum_{n=0}^\infty e^{-\frac{16 n\pi^2}{\nu \lambda}}\ .
\ee
Writing this in terms of the  dynamical scale $\Lambda_{\rm eff}$,
one obtains an explicit expression for the OPE to all orders,
\be
\label{factorizL}
\frac{\partial F_w}{\partial\alpha}=
\left( 2\Lambda_{\rm eff}^2 -\Lambda_{\rm eff}^4\left(\frac{1}{m^2}+\frac{1}
{M^2}\right)\right)\sum_{n=0}^\infty 
\left(\frac{\Lambda_{\rm eff}^2}{m M}\right)^{2n}
\ .
\ee

\medskip

\paragraph{Remarks on integrated correlators:}
Recently, some observables of $\mathcal{N}=4$ superconformal field theories representing integrated correlation functions have been determined exactly using localization. 
One of these correlation functions correspond to the integrated correlator of four primaries of the stress tensor supermultiplet of the ${\cal N} = 4$ theory \cite{Binder:2019jwn,Dorigoni:2021guq}. It is obtained from the mass-deformed ${\cal N} = 2^*$ theory by
differentiating $\log Z$,
\be\label{eq: mixed derivative}
 4\tau_2^2\partial_\tau \partial_{\bar\tau} \partial_m^2 \log Z\ \bigg|_{m=0}
\ee
where $\tau $ represents the Yang-Mills coupling. This can be identified with an integral over the insertion points
in the four-point correlator
(see also \cite{Alday:2023pet,Dorigoni:2022cua,Chester:2020vyz,Chester:2020dja} for related research).  

In $\cN=2$ theories, a notable relation between integrated correlators and derivatives of the free energy was explored in
\cite{Chester:2022sqb}. These involve integrated correlation functions 
of the flavor conserved
current multiplet, having the moment map operator as lowest component. The correlation function is related to
\be
\label{mmmm}
\partial^4_m \log Z_{\cN=2}[m]\bigg|_{m=0}\ .
\ee
This observable has been further investigated  in $\cN=2$ SYM theories with matter fields in diverse representations \cite{Fiol:2023cml,Billo:2023kak,Billo:2024ftq}.
The  analog of \eqref{eq: mixed derivative} for $\cN=2$ theories has only been studied for 
{\bf E} theory and represents a mixed integrated correlator among two
Coulomb-branch operators of dimension 2 and two moment-map operators \cite{Billo:2023kak}.
The precise relation between \eqref{eq: mixed derivative} and integrated correlators in more general $\cN=2$ superconformal theories remains to be fully established.

In this section, we focus in the non-conformal setting of massive  SQCD with $N_f<2N$. 
As noted above, 
the SQCD case with $N_f<2N$ is obtained from the theory with two mass scales $m,\ M$ by taking $M=\infty$ with $\Lambda =M e^{-\frac{\alpha}{2 \zeta_M}}$ fixed. 
In contrast to the $\cN=4$ theory and to $\cN=2$ superconformal theories, where the integration measure and part of the kinematical structure are determined by conformal invariance and supersymmetry, in the present non-conformal example the relation between derivatives of the
free energy and integrated correlators is less well-defined.
Finding the precise structure of integrated correlators for $N_f<2N$ SQCD theory,
which is an asymptotically free theory with a running coupling, is  challenging. 
Nonetheless,  derivatives of the free energy still represent observables, which must correspond to insertions of operators in the path integral. We will find that these observables exhibit intriguing properties.

The perturbative part of the partition function depends only on $\tau_2$, which, in the large $N$ limit, is traded by $1/\lambda$. Thus, modulo a multiplicative constant, the analog of \eqref{eq: mixed derivative} and \eqref{mmmm} are 
\be
{\cal J}\equiv \partial_\alpha^2 \partial_m^2\, F\ ,\qquad {\cal R}\equiv  \partial_m^4\, F\ .
\ee
Here we examine the properties of ${\cal J}$ 
 at finite mass, anticipating that this observable will exhibit an interesting feature.
 Our starting point is the equation \eqref{hesta},
which reads
\be\label{fgy}
 \frac{\partial^2 F}{\partial\alpha^2 }=-\frac{\mu^2}{2}\ .
\ee
where have set $\nu =1$
(a similar calculation applies to the {\bf B}$^*$ and {\bf C}$^*$ theories upon setting $\nu=1/2$).

Let us first consider the weak-coupling phase.
For the general two-mass scale theory,
 $\mu$ is determined by \eqref{mueq}.
In the above $M=\infty $ limit,
this becomes ($\zeta\equiv\zeta_m$) 
\be\label{eq: mu2 SQCD}
\frac12 \log\mu^2 -\zeta \log\left(m+\sqrt{m^2-\mu^2}\right)=(1-\zeta)\,\log(2\Lambda)\ .
\ee
A calculation gives
\be\label{eq: J weak}
{\cal J}_{\rm weak}=\left(2\zeta-1\right)\frac{1-u^2}{(\zeta+u(1-\zeta))^3} \,  ,
\ee
where
\be
\label{uuf}
u\equiv \sqrt{1-\frac{\mu^2}{m^2}} \quad , \quad u\in [0,1] \,\, .
\ee
Remarkably, ${\cal J}_{\rm weak}$ identically vanishes for $\zeta=1/2$.
 We recall that $\zeta=1/2$ corresponds
to a number of flavors satisfying
 $\frac{N_f}{N}\to 1$ for $N\to\infty$.

The critical point is approached as $u\to 0$ and yields
\be
\label{criJ}
{\cal J}_{\rm weak} \,\, \xrightarrow{u\to 0} \,\, \frac{2\zeta-1}{\zeta^2 }\ .
\ee
On the other hand, in the strong coupling phase, one has 
$\mu=2\Lambda $. Therefore, in this phase \eqref{fgy} is independent of $m$. One obtains
\be
{\cal J}_{\rm strong}= 0\ .
\ee
Comparing with \eqref{criJ}, we find that  ${\cal J}$    is discontinuous across the phase transition for any $\zeta\neq \tfrac12$, while it is continuous when  $\zeta=\tfrac12$, as in this case 
${\cal J}_{\rm weak}= {\cal J}_{\rm strong}= 0$ for any coupling $\Lambda$.

\smallskip

We now turn to the calculation of  ${\cal R}=\partial^4_m F$. 
In the previous formulas \eqref{freedebil} and \eqref{freestrong} we have computed $\partial_\alpha F$, which is much simpler to evaluate than the evaluation of $F$ directly.
Integrating $\partial_\alpha F$ gives a formula for $F$ modulo the addition of an arbitrary function of $m$.

A complete formula for the free energy, including its mass dependence, has not yet been obtained in  $\cN =2$ SQCD with $N_f<2N$ massive hypermultiplets. However, computing 
$\partial^4_m F$
 requires accounting for the full mass dependence.

To proceed, we first integrate \eqref{freestrong} in $\alpha$. We get
\be
F_s=\zeta \alpha  m^2- 2(1-\zeta)^2 M^2 \log\left(1+e^{-\frac{\alpha}{1-\zeta}}\right)+h(m,M)\ . 
\ee
In the   $M=\infty$ limit with $\Lambda =M e^{-\frac{\alpha}{2 (1-\zeta)}}$ fixed, the free energy becomes 
\be
\label{FFhh}
F_s=-2\zeta(1-\zeta)   m^2\log\frac{2\Lambda}{m} - 2(1-\zeta)^2  (\Lambda^2-\frac{m^2}{4}) +h_s(m)\ , 
\ee
where $h(m,M)$ has been redefined to absorb a $\log M$ divergent term and to have $F_s=h_s(m)$ at the transition point $m=2\Lambda$. 
On the other hand, integrating \eqref{freedebil} in $\alpha$
and taking the same limit, gives the formula
\be\label{SQCDfree}
F_w=-\frac{m^2}{2}\left\{(1-\zeta)(1-u)\big[1+5\zeta+(1-\zeta)u]-(1-\zeta ) (1+5 \zeta ) +4 \zeta(1-2\zeta)\log(1+u)\right\}+h_w(m)\ ,
\ee
where the  dependence on the coupling $\Lambda $ is incorporated in $u$.
Explicitly, using \eqref{uuf} and \eqref{eq: mu2 SQCD} one gets
$$
(1+u)^{1-2\zeta}(1-u)= \left(\frac{2\Lambda}{m}\right)^{2-2\zeta}\ .
$$
The free energy must be continuous at the phase transition at $m=2\Lambda$ (where $u=0$). This sets $h_s(m)=h_w(m)$.

We now determine $h_s(m)$.
 The starting point is the formula
\be
 \frac{ \partial F}{\partial m} = \zeta \int_{-\mu}^\mu dx\, \rho(x)\left(-K(x+m)+K(x-m)\right)\ ,
\ee
where the densities in the weak and strong coupling regime
are given by \eqref{mumenor} and \eqref{mumayor} (with $M\to m$).
For our purpose it is sufficient to consider the strong coupling regime, with the density given in \eqref{mumayor} and $\mu=2\Lambda $. We first compute the contribution from the delta functions. This gives
\be
-\zeta^2 K(2m)\approx -\zeta^2 2m \log (4m^2)
\ee
where the RHS applies in the decompactification limit
(and we neglected a regularization dependent linear term $\gamma\, m$, as it will not contribute to $\partial^4_m F$).
In the decompactification limit the remaining contribution is
\be
\int_{-\mu}^\mu dx\, \frac{\zeta(1-\zeta)}{\pi\sqrt{\mu^2-x^2}}\, \left(-(x+m)\log(x+m)^2+(x-m)\log(x-m)^2\right) =-2m\, \zeta\, (1-\zeta)\, \big(\log\Lambda^2+2\big)\ ,\ \  \mu>m\ .
\ee
%
%
Thus we find 
\be
\label{aaaa}
\partial_m F_s = -\zeta^2 2m \log (4m^2)-4m\, \zeta\, (1-\zeta)\, \log\Lambda\ ,
\ee
where again we have ignored a coupling-independent  linear term in $m$.
As explained, the linear  term in $m$ is regularization dependent and does not contribute to $\partial^4_m F$ (it can be absorbed into a redefinition of $\Lambda $ by a rescaling).

On the other hand, differentiating \eqref{FFhh} with respect to $m$ gives
\be
\label{bbbb}
\partial_m F_s = \zeta(1-\zeta) 2m \log (m^2)+h_s'(m)-4m\, \zeta\, (1-\zeta)\, \log\Lambda\ .
\ee
Comparing \eqref{aaaa} and \eqref{bbbb}, we find
\be
h_s'(m)=-2 \zeta  m \log \left(m^2\right)\ ,\qquad h_s(m)=-\zeta m^2 \log
   \left(m^2\right)\ .
\ee

We  have now derived a complete expression for the free energy in both the weak and strong coupling phases, including the  mass-dependent terms. We proceed by calculating
 ${\cal R}_s$ and ${\cal R}_w$. To begin with, we note that there is a discontinuity in the third derivative of the mass at the critical point,  
\be
(\partial_m^3 F_s-\partial_m^3 F_w)_{\Lambda=\frac{m}{2}}=\frac{4 (1-\zeta)^2}{m}\ .
\ee
For  ${\cal R}_s$ and ${\cal R}_w$ we find
\bea
&&{\cal R}_s=\partial_m^4F_s =\frac{4 \zeta^2}{m^2}\ ,
\\
\nonumber\\
&&{\cal R}_w=\partial_m^4F_w =\frac{4 \zeta}{m^2 (\zeta -\zeta  u+u)^3}  \left(
A+Bu+Cu^2+Du^3\right)\ ,
\eea
where
\bea
&& A=1-4 \zeta +6 \zeta ^2-4 \zeta
   ^3+2 \zeta ^4\ ,\ \ B=2 \zeta (1-\zeta)   \left(3
   \zeta ^2-2 \zeta +1\right)\ ,\ \ 
   \nonumber\\
   &&C= 6 \zeta ^2(1-\zeta)^2  \ ,\ \ D=2 (1-\zeta)^3 \zeta \ .
\eea
It is interesting to note that ${\cal R}_s$ does not depend on the coupling $\Lambda$.
On the other hand, ${\cal R}_w$ depends on the coupling through $u$, but, strikingly, the dependence
on $\Lambda$ cancels out precisely in the particular case $\zeta=\tfrac12$, where
\be
{\cal R}_w\bigg|_{\zeta=\tfrac{1}{2}}=\frac{2}{m^2}\ .
\ee
Thus, we observe that $\partial^4_m F$
 also exhibits special properties at $\zeta=\tfrac12$: as $\Lambda$ is increased from 0 up to the critical value $\Lambda_c=m/2$, $\partial^4_m F$ is constant; it exhibits a finite jump
 from $2/m^2$ to $1/m^2$ in crossing the critical point, and it remains constant all the way from $\Lambda_c$ to infinity. For any other generic value, $0<\zeta<1$, ${\cal R}_w$ exhibits a non-trivial 
dependence on the coupling $\Lambda $.
 Elucidating the general features of SQCD at $\zeta=\tfrac{1}{2}$ remains an important and compelling direction for further investigation.

\bigskip

\medskip

\paragraph{Conclusions:} $\cN =2$ theories deformed by mass terms for the hypermultiplets
exhibit phase transitions at  critical couplings where the width of the eigenvalue distribution reaches one of the mass parameters. At these critical points new massless excitations  contribute 
to the saddle point. The mechanism 
is generic and we have confirmed that it also applies to other examples of $\cN=2$ theories with different hypermultiplet representations.
In some exceptional cases, a massive  $\cN =2$ theory may not undergo any phase transition, because the critical point where $\mu$ reaches a mass parameter arises at infinite coupling. An example is
SQCD with $N_f=2N$ fundamental hypermultiplets of mass $M$.

We have explored a theory characterized by two mass scales, $m$ and $M$, which describes a flow from either the {\bf B} or  {\bf C}  superconformal theories ($\nu=1/2$) --~or from superconformal QCD ($\nu=1$)~-- to pure $\cN =2$ SYM.
The theory exhibits a number of features of interest. 
At weak coupling the free energy
is given by a power expansion in the effective scale $\Lambda_{\rm eff}$ given in \eqref{lamef}.
When the 't Hooft coupling $\lambda $ is increased, the eigenvalue distribution expands until the width $\mu$ meets $m$.
Across this point, the free energy has a discontinuity in the third derivative with respect to $\lambda $.
At strong coupling, the free energy becomes an expansion in
integer powers of $1/\lambda$.
In the limit of infinite coupling, $\mu$ asymptotically approaches the second mass scale $M$.
The expressions simplify when $\zeta_m=\zeta_M$ (i.e. same number of multiplets with mass $m$ and mass $M$). In this case the strong coupling expansion is simply obtained by expanding $\tanh(\alpha/\nu)$ in \eqref{finalx2} in powers of $\alpha$ (with coefficients  given in terms of Bernoulli numbers).

The two-scale model clarifies why the single mass deformation of superconformal QCD does not undergo any phase transition. In the limit $m\to M$, the critical coupling given in \eqref{lambdacr} tends to infinity.
The phase transition emerges for an arbitrary small mass scale separation, since $\lambda_c$ is finite
for any $m<M$.

Some open problems include understanding
the physics of the OPE given in \eqref{factoriz} and \eqref{opex}, the critical properties of more general integrated correlators and the identification of the underlying dynamics behind the special properties of the theory at $\zeta_m=\tfrac{1}{2}$.

\bigskip\bigskip

We acknowledge financial support from grant 2021-SGR-249 (Generalitat de Catalunya) and  by the Spanish  MCIN/AEI/10.13039/501100011033 grant PID2022-126224NB-C21.

\bigskip

\bigskip



\bibliographystyle{JHEP}
\bibliography{N=2massV2.bib} 



\end{document}